\documentclass{article}

\usepackage{arxiv}

\usepackage{siunitx}
\usepackage{bm}
\usepackage{algorithm}
\usepackage{graphicx}
\usepackage{algpseudocode}
\usepackage{physics}
\usepackage{amsmath}
\usepackage{amssymb}
\usepackage[table]{xcolor}
\definecolor{royal_blue}{rgb}{0.255, 0.412, 1}
\definecolor{medium_blue}{rgb}{0, 0, 0.804}
\usepackage{hyperref}
\hypersetup{colorlinks=true,linkcolor=medium_blue, citecolor=medium_blue}
\usepackage{subcaption} 
\usepackage{multirow}
\usepackage{float}
\usepackage{layouts}

\title{The Radial Point Interpolation Mixed Collocation (RPIMC) Method for the Solution of Transient Diffusion Problems}

\author{
  Konstantinos A. ~Mountris\thanks{[mail] kmountris@unizar.es \quad [url] https://www.mountris.org} \\
  Arag\'on Institute of Engineering Research, IIS Arag\'on\\
  University of Zaragoza\\
  Spain, Zaragoza, ZGZ 50018 \\
  \texttt{kmountirs@unizar.es} \\
  \AND
  Esther ~Pueyo \\
  Arag\'on Institute of Engineering Research, IIS Arag\'on, \\
  CIBER-BBN\\
  University of Zaragoza\\
  Spain, Zaragoza, ZGZ 50018 \\
  \texttt{epueyo@unizar.es} \\
}

\begin{document}
\maketitle

\begin{abstract}
The Radial Point Interpolation Mixed Collocation (RPIMC) method is proposed in this paper for transient analysis of diffusion problems. RPIMC is an efficient purely meshfree method where the solution of the field variable is obtained through collocation. The field function and its gradient are both interpolated (mixed collocation approach) leading to reduced $C$-continuity requirement compared to strong-form collocation schemes. The accuracy of RPIMC was evaluated in heat conduction benchmark problems and compared against the Meshless Local Petrov-Galerkin Mixed Collocation (MLPG-MC) method and the Finite Element Method (FEM). These three methods were subsequently applied to solve electrical propagation in a cardiac tissue slab and a biventricular geometry. In benchmark problems, RPIMC achieved high accuracy, of the same order as FEM, and always higher than MLPG-MC due to the delta Kronecker property of RPIMC. In solving cardiac electrical propagation, RPIMC rendered activation time maps similar to those of FEM, with an improved performance compared to MLPG-MC. As a conclusion, RPIMC is shown to be a promising meshfree alternative to FEM for transient diffusion problems.
\end{abstract}


\newpage

\section{Introduction} \label{sec:intro}

The diffusion equation describes physical phenomena where motion is driven by the gradient of the field variable. Time-dependent problems like heat and mass transport \cite{Incropera2013}, unsteady viscous fluid flow \cite{Patankar1983} and magneto-hydrodynamics flow \cite{Hughes1994} can be solved by the transient diffusion equation. Also, the diffusion equation appears in the description of coupled phenomena, such as the transport of chemical or biological reactions by diffusive propagation in a medium (reaction-diffusion phenomena) \cite{Britton1986}. Mathematically, reaction-diffusion problems are described by a coupled set of ordinary differential equations (ODEs) that describes the reactive term and a partial differential equation (PDE) that describes the diffusive term. Usually, a much smaller time scale is required for the reactive term than for the diffusive term and operator splitting techniques are used to decouple the problem and compute a numerical solution efficiently \cite{Strang1968,McLachlan2002}. Among the various available methods to solve the PDE of the diffusive term, of great interest are Meshless Methods (MMs).

MMs, in contrast to mesh-based methods, do not require connectivity information for the construction of basis functions. Therefore, domains with irregular geometry, nonlinearity and discontinuity can be treated efficiently. The use of MMs to solve both the steady and transient diffusion equation has been extensively reported. Steady-state heat conduction in isotropic and functionally graded materials has been solved successfully by the Meshless Point Collocation (MPC) method \cite{Skouras2011}. In \cite{Sarler2006}, an explicit collocation method with local Radial Basis Functions (RBFs) has been successfully applied to solve the transient diffusion equation in two-dimensional (2D) domains. The collocation methods have demonstrated high efficiency due to the compact support and the small bandwidth in linear algebraic systems \cite{bourantas2018strong}. However, accuracy has been shown to deteriorate near the Neumann boundaries due to the requirement for the approximation of spatial derivatives, which is significantly less accurate than the approximation of the field variable \cite{Libre2008}. 

On the other hand, in the Element Free Galerkin (EFG) meshless method \cite{Belytschko1994}, which is based on the Galerkin weak formulation, the Neumann boundary conditions (BCs) are satisfied naturally, similarly to the Finite Element Method (FEM). In EFG, the notion of a background mesh is introduced for the generation of quadrature points and the evaluation of the weak form's spatial integrals. Application of EFG for the solution of heat transfer problems has been rigorously explored \cite{Singh2003,Singh2006,Yang2010}. Moreover, the improved Moving Least Squares (MLS) approximants have been proposed to enhance the handling of Dirichlet BCs and the efficiency of EFG for three-dimensional (3D) heat conduction problems \cite{Zhang2013}. Maximum entropy approximants that possess the weak-Kronecker delta property for direct imposition of Dirichlet BCs have also been proposed in the framework of EFG \cite{arroyo2006,millan2015,mountris2019,wittek2020meshless}.

A different approach is considered in the Meshless Local Petrov-Galerkin (MLPG) method \cite{Atluri2004a,Li2008,Han2005}, where quadrature points are generated in individual local quadrature domains centered at each field node and the trial and test functions can be selected from different spaces. The flexibility in the selection of the test functions offers the possibility to construct different variations of the MLPG method \cite{Atluri2005}. By choosing the Dirac function as the test function and interpolating both the field function and its gradient, the MLPG Mixed Collocation (MLPG-MC) method is derived \cite{Atluri2006}. MLPG-MC has minimum computational cost since no integration is performed. Compared to standard collocation methods, MLPG-MC demonstrates reduced deterioration at the Neumann boundaries, as the order of the spatial derivatives is reduced through the interpolation of the field function's gradient. The MLPG-MC method has been successfully applied to solve inverse Cauchy problems for steady-state heat transfer \cite{Zhang2014}.

Variations of the MLPG method using different trial functions have been investigated extensively \cite{Liu2005,Liu2007}. Since MLS basis functions do not possess the delta Dirac property, special treatment to impose the Dirichlet BC is required. To address this issue, the Local Radial Point Interpolation method has been proposed \cite{Liu2003}, in which the MLS basis functions are replaced with Radial Point Interpolation (RPI). The RPI basis functions possess the Kronecker delta property and Dirichlet BC imposition is straightforward as in FEM and maximum entropy approximants. The Local RPI method has been used to successfully solve problems in free vibration analysis \cite{Liu2001}, incompressible flow \cite{Liu2003}, material non-linearity \cite{Gu2007} and transient heat conduction \cite{Wang2013}, among others. However, to our knowledge, RPI has not been so far evaluated in the mixed collocation variant.

The purpose of the present study is to investigate the performance of the mixed collocation method using the RPI basis functions for the solution of transient diffusion problems. The method is subsequently applied to solve the monodomain reaction-diffusion equation for electrical impulse propagation in the heart \cite{Potse2006}. The standard approach to solve the monodomain model involves the use of the operator splitting method to decouple the reaction and diffusion parts and solve them separately. It is for that reason that in this study pure transient diffusion problems are initially considered. Without loss of generality, the method is evaluated in 2D and 3D benchmark problems of transient heat conduction. The structure of the paper is the following. In section \ref{sec:rpi_theory}, the theory of the RPI basis functions is reviewed. In section \ref{sec:rpimc}, the mathematical formulation and implementation details of the Radial Point Interpolation Mixed Collocation (RPIMC) method are presented. In section \ref{sec:benchmark}, the RPIMC method is first evaluated in 2D and 3D heat conduction benchmark problems and subsequently applied to solve the monodomain model for a 3D tissue slab and a biventricular geometry. The time efficiency of RPIMC method for the solution of the monodomain model is profiled. Finally, in section \ref{sec:conclusions} some concluding remarks are provided.

\section{Radial point interpolation review} \label{sec:rpi_theory}

In RPI \cite{wang2002point}, RBFs augmented with polynomials are used to approximate the field function. In contrast to MLS, RPI possesses the Kronecker delta property, therefore essential boundary conditions are imposed directly. For any field function $u(\bm{x})$, defined in the domain $\Omega \subset \mathbb{R}^d$, the RPI approximation $u^h(\bm{x_I})$ at a point of interest $\bm{x_I} \in \mathbb{R}^d$ is given by:
\begin{equation} \label{eq:rpi}
    u^h(\bm{x_I}) = \sum_{i=1}^n r_i(\bm{x_I})a_i(\bm{x_I}) + \sum_{j=1}^m p_j(\bm{x_I})b_j(\bm{x_I})  = \bm{r}^T(\bm{x_I}) \bm{a}(\bm{x_I}) + \bm{p}^T(\bm{x_I}) \bm{b}(\bm{x_I})
\end{equation}
\noindent where $r_i(\bm{x_I})$ are the RBFs and $p_j(\bm{x_I})$ are the polynomial basis functions, $a_i(\bm{x_I})$ and $b_j(\bm{x_I})$ denote the corresponding coefficients, $n$ is the number of neighbor nodes in the local support domain of $\bm{x_I}$, and $m$ is the number of polynomial terms. In Equation (\ref{eq:rpi}), different forms of RBFs can be used to represent $r_i(\bm{x_I})$. In this study, we used the Multi-Quadric RBFs (MQ-RBFs) due to their satisfactory performance reported in previous studies \cite{Liu2004,Liu2007}. In 2D, the MQ-RBFs are given by:
\begin{equation} \label{eq:mq_rbf}
    r_i(\bm{x_I}) = \left(d_{Ii}^2+r_c^2\right)^q = \left[ (x_I-x_i)^2 + (y_I-y_i)^2 + r_c^2  \right]^q
\end{equation}
\noindent where $r_c$ and $q$ are positive-valued shape parameters of the MQ-RBF and $d_{Ii}$ is the Euclidean norm between the point of interest $\bm{x_I}=(x_I, y_I)$ and the $i^{th}$ neighbor node $\bm{x_i}=(x_i, y_i)$. Analogous MQ-RBFs are defined in 3D. Rectangular and cuboid local support domains for 2D and 3D problems, respectively, were constructed in this study. Following the notation in \cite{Liu2004}, the shape parameter $r_c$ is given by:
\begin{equation} \label{eq:rpi_param_c}
    r_c = \alpha_c d_c
\end{equation}
\noindent where $\alpha_c$ is a dimensionless constant and $d_c$ denotes the average nodal spacing in the proximity of the point of interest $\bm{x}$. The effect of the choice of $\alpha_c$ and $q$ on the approximation accuracy has been investigated in \cite{Liu2002, Liu2003}. In this study, parameter values $\alpha_c = 1.5$ and $q=1.03$ were used in all the problems of section \ref{sec:benchmark}.

The $k^{th}$ order polynomial basis function $\bm{p}(\bm{x})$ in Equation (\ref{eq:rpi}) is given, in 2D (analogously for 3D), by:
\begin{equation} \label{eq:poly_basis}
    \bm{p}(\bm{x}) = \bm{p}(x,y) = \{ 1,x,y,xy,x^2,y^2,\dotso,x^k,y^k \}^T.
\end{equation}
In this work, we used the linear polynomial basis functions $(k=1)$. The coefficients $a_i(\bm{x_I})$, $b_j(\bm{x_I})$ are obtained by requiring the field function to pass through all the $n$ field nodes in the local support domain, expressed in matrix form:
\begin{equation} \label{eq:field_function_rpi_mat}
    \bm{u_s} = \bm{R} \bm{a}(\bm{x_I}) + \bm{P} \bm{b}(\bm{x_I})
\end{equation}
\noindent where $\bm{u}_s = \{u_1, u_2, \dotso, u_n\}^T$ is the vector of the field function parameters at the nodes of the local support domain, $\bm{R}$ is the RBF moment matrix of size $n \times n$, and $\bm{P}$ is the polynomial moment matrix of size $n \times m$. A unique solution to Equation (\ref{eq:field_function_rpi_mat}) is obtained by applying the following constraint condition \cite{Golberg1999}:
\begin{equation} \label{eq:constraint_cond}
    \bm{P}^T \bm{a}(\bm{x_I}) \equiv \sum_{i=1}^n p_j(\bm{x}_i) a_i(\bm{x_I}) = 0, \quad j=1,2,\dotso,m.
\end{equation}
By combining Equations (\ref{eq:field_function_rpi_mat}) and (\ref{eq:constraint_cond}) the following equations are obtained: 
\begin{equation} \label{eq:field_function_rpi_augment}
    \bm{\Tilde{u}_s} = \begin{bmatrix}
                    \bm{u_s} \\
                    \bm{0}
                  \end{bmatrix} = \begin{bmatrix}
                                        \bm{R} & \bm{P} \\
                                        \bm{P}^T & \bm{0}
                                  \end{bmatrix} \begin{bmatrix}
                                                    \bm{a}(\bm{x_I}) \\
                                                    \bm{b}(\bm{x_I})
                                                \end{bmatrix} = \bm{G} \bm{a_0}(\bm{x_I})
\end{equation}
\noindent and the unique solution is given by:
\begin{equation} \label{eq:inv_G}
    \bm{a_0}(\bm{x_I}) = \begin{Bmatrix}
                \bm{a}(\bm{x_I}) \\
                \bm{b}(\bm{x_I})
                \end{Bmatrix} = \bm{G}^{-1} \bm{\Tilde{u}_s} .
\end{equation}
To ensure that $\bm{G}^{-1}$ is not singular, $\bm{R}^{-1}$ should exist. The existence requirement is usually satisfied, even for arbitrarily scattered nodes \cite{Hardy1990,Wendland1998}, rendering RPI a stable approximation method. Finally, the RPI approximation $u^h(\bm{x_I})$ at $\bm{x_I}$ as a function of the RPI basis function:
\begin{equation} \label{eq:rpi_shape}
    \bm{\phi}(\bm{x_I}) = \{ \phi_1(\bm{x_I}) \quad \phi_2(\bm{x_I}) \quad \dotso \quad \phi_n(\bm{x_I}) \}^T
\end{equation}
\noindent is obtained from Equations (\ref{eq:rpi}) and (\ref{eq:inv_G}) as follows:
\begin{equation} \label{eq:rpi_approx}
\begin{split}
    u^h(\bm{x_I}) = 
        \begin{Bmatrix}
            \bm{r}^T(\bm{x_I}) \hspace{5pt} \bm{p}^T(\bm{x_I}) 
        \end{Bmatrix} \bm{G}^{-1} \bm{\Tilde{u}_s} =      
        \bm{\phi}^T(\bm{x_I}) \bm{u_s} = \sum_{i=1}^n \phi_i(\bm{x_I}) u_i .
\end{split}
\end{equation}
The derivatives of $u^h(\bm{x})$ can be computed as:
\begin{equation} \label{eq:rpi_deriv}
    u^h_{,J}(\bm{x_I}) = \bm{\phi}_{,J}^T(\bm{x_I}) \bm{u_s} .
\end{equation}
\noindent where $J$ denotes spatial coordinate and the comma symbol designates partial differentiation with respect to $J$.

\section{Radial Point Interpolation Mixed Collocation Method} \label{sec:rpimc}

In this section the theoretical aspects of RPIMC and its computer implementation are described. The RPIMC theoretical formulation is based on the principles of MLPG-MC \cite{Zhang2014,Atluri2006}. However, the RPI basis function is used as trial function instead of the MLS. The Dirac function is selected as the test function, thus reducing the integration over local domains to collocation. Without loss of generality, the field variable $u$ is used to represent the temperature field in the following.

\subsection{Theoretical aspects} \label{subsec:rpimc_theory}

Let's consider the balance equation of heat transfer in a domain $\Omega$ with boundary $\partial{\Omega} = \partial{\Omega_u} \cup \partial{\Omega_q}$ given by:
\begin{align}
    c \rho \pdv{u(\bm{x},t)}{t} + \div{\bm{q}(\bm{x},t)}  & = f(\bm{x}, t)          \quad  \text{in } \Omega  \label{eq:heat_balance} \\
    u(\bm{x},t)             & = \Bar{u}(\bm{x},t)    \quad  \text{at } \partial{\Omega_u} \label{eq:dirichlet} \\
    -\bm{n} \cdot k \grad{u}(\bm{x},t)  & = \Bar{q}(\bm{x},t)  \quad  \text{at } \partial{\Omega_q} \label{eq:neumann}
\end{align}
\noindent where $c$ is the specific heat capacity, $\rho$ is the material density, $u(\bm{x},t)$ is the temperature field, $\bm{q}(\bm{x},t)$ is the heat flux, $f(\bm{x},t)$ denotes the sum of any heat sources acting in the domain $\Omega$, $\Bar{u}$ is the prescribed value of the temperature field on the Dirichlet boundary $\partial{\Omega_u}$, $\Bar{q}$ is the prescribed value of the heat flux on the Neumann boundary $\partial{\Omega_q}$, $\bm{n}$ is the outward unit vector normal to $\partial{\Omega_q}$ and $k$ is the thermal diffusivity coefficient. 

The heat flux $\bm{q}$ can be expressed by the heat flux - temperature gradient relation as:
\begin{equation} \label{eq:flux_temp}
    \bm{q}(\bm{x},t) = -k \grad{u}(\bm{x},t).
\end{equation}
\noindent The RPI basis functions are used to interpolate the temperature field $u(\bm{x},t)$ and the heat flux fields $q_J(\bm{x},t)$, thus obtaining:
\begin{align}
    u(\bm{x},t)   & = \sum_{i=1}^n \phi^i(\bm{x}) u^i(t)   \label{eq:nodal_temp} \\
    q_J(\bm{x},t) & = \sum_{i=1}^n \phi^i(\bm{x}) q_J^i(t) \label{eq:nodal_flux}
\end{align}
\noindent where $n$ is the number of field nodes in the local support domain of $\bm{x}$.

From Equations (\ref{eq:flux_temp}) and (\ref{eq:nodal_temp}), the relationship of the nodal heat fluxes to nodal temperatures is obtained. The heat flux-to-temperature relationship at a node $\bm{x_I}$ for time $t$ is given by:
\begin{equation} \label{eq:flux_temp_nodal}
    \bm{q}(\bm{x_I},t) = -k \sum_{i=1}^n \grad{\phi^i}(\bm{x_I}) u^i(t), \quad I = 1, 2, \dotso , N
\end{equation}
\noindent where $N$ is the number of field nodes in the discretization of the domain $\Omega$. In matrix form, the heat flux to temperature relationship is written as:
\begin{equation} \label{eq:flux_temp_nodal_matrix}
\bm{q} = \bm{K_a u}
\end{equation}
\noindent where $\bm{q}$ is a vector containing the heat fluxes at each field node $\bm{x_I}$, $\bm{K_a}$ is a sparse matrix containing the partial derivatives of the RPI basis functions at the nodes in the local support of each field node scaled by $(-k)$ and $\bm{u}$ is a time-dependent vector containing the nodal temperature values.

Introducing Equations (\ref{eq:nodal_temp}-\ref{eq:flux_temp_nodal}) in Equation (\ref{eq:heat_balance}), the RPIMC formulation for a node $\bm{x_I}$ at time $t$ is given by:
\begin{equation} \label{eq:heat_balance_nodal}
    c \rho \sum_{i=1}^n \phi^i(\bm{x_I}) \pdv{u^i(t)}{t} + \sum_{i=1}^n \grad{\phi^i}(\bm{x_I}) \bm{q}^i(t) = f(\bm{x_I}, t), \; I = 1, 2, \dotso , N
\end{equation}
which can be expressed in terms of the temperature field as:
\begin{equation} \label{eq:heat_balance_nodal_T}
    c \rho \sum_{i=1}^n \phi^i(\bm{x_I}) \pdv{u^i(t)}{t} - k \sum_{i=1}^n \div{(\grad{\phi^i}(\bm{x_I})u^i(t))} = f(\bm{x_I}, t), \; I = 1, 2, \dotso , N
\end{equation}
\noindent Equations (\ref{eq:heat_balance_nodal}) and (\ref{eq:heat_balance_nodal_T}) can be written in the equivalent matrix form as:
\begin{equation} \label{eq:heat_balance_nodal_mat}
    \bm{M} \dot{\bm{u}} + \bm{K_s q} = \bm{f} 
\end{equation} 
\begin{equation}
    \bm{M} \dot{\bm{u}} + \bm{K u} = \bm{f}, \quad \bm{K} = \bm{K_s K_a}
\end{equation}
where $\bm{f}$ is a $N \times 1$ time-dependent vector containing the values of $f(\bm{x_I}, t)$ for all field nodes $\bm{x_I}$, $I = 1, 2, \dotso , N$.

\subsection{Boundary conditions imposition} \label{subsec:rpimc_bc}
To impose Dirichlet BCs (Equation (\ref{eq:dirichlet})) in the mixed collocation method, the prescribed temperature values at field nodes $\bm{x_I}$ belonging to the Dirichlet boundary $\partial \Omega_u$ are considered. The collocation method is used to enforce them:
\begin{equation} \label{eq:dirichlet_collocation}
    \sum_{i=1}^n \phi^i(\bm{x_I}) u^i(t) = \Bar{u}(\bm{x_I},t),
\end{equation}
\noindent where $n$ is the number of field nodes in the support domain of $\bm{x_I}$. Due to the Kronecker delta property of the RPI basis functions, Equation (\ref{eq:dirichlet_collocation}) is reduced to strong imposition in the RPIMC method and Dirichlet BCs are satisfied exactly. This is in contrast to the MLPG-MC, in which special treatment for the Dirichlet BCs is required due to the lack of the Kronecker delta property of the MLS basis functions.

Neumann BCs (Equation (\ref{eq:neumann})) are enforced using the penalty method described in \cite{Overvelde2012}. The rows of the matrices $\bm{K}_s$ and $\bm{K}_a$ are reordered such that $\bm{K}_s^T = [\bm{K}_s^1 \quad \bm{K}_s^2 ]$ and $\bm{K}_a^T = [\bm{K}_a^1 \quad \bm{K}_a^2 ]$. Superscript 1 denotes the $\gamma_r$ nodes on the $\partial \Omega_q$ boundary (Neumann nodes) and superscript 2 denotes the $\gamma_u$ nodes on the $\partial \Omega_u$ boundary (Dirichlet nodes). $\gamma_{in}$ denotes the nodes in the interior of the domain $\Omega$, such that the total number of nodes is $N = \gamma_r + \gamma_u + \gamma_{in}$. For a given time $t$, the matrix form of Equation (\ref{eq:neumann}) is given by:
\begin{equation} \label{eq:flux_bc_matrix}
    \bm{N_r} \bm{q}^1 = \bm{\Bar{q}_r},    
\end{equation}
where $\bm{q}^1$ is the vector of the nodal heat fluxes for the $\gamma_r$ nodes. $\bm{N_r}$ is the matrix containing the normal vectors and $\bm{\Bar{q}_r}$ is the vector of the prescribed heat fluxes for the $\gamma_r$ nodes, given by:
\begin{equation}
    \bm{N_r} = \begin{bmatrix}
                    \bm{n}^1 & & 0\\
                    & \ddots & \\
                    0 & & \bm{n}^{\gamma_r}
                \end{bmatrix}
    \quad \text{and} \quad
    \bm{\Bar{q}_r} = \begin{bmatrix}
                        \Bar{q}^1 \\
                        \vdots \\
                        \Bar{q}^{\gamma_r} \\
                        \end{bmatrix}.
\end{equation}
The Neumann BCs are enforced at the nodes $\gamma_r$ by multiplying Equation (\ref{eq:flux_bc_matrix}) with the penalty factor $\alpha \bm{N_r}^T$ and adding it to Equation (\ref{eq:flux_temp_nodal_matrix}) to obtain:
\begin{equation} \label{eq:modified_traction}
    \bm{q}^1 + \alpha \bm{N_r}^T \bm{N_r} \bm{q}^1 = \bm{K_a}^1 \bm{u} + \alpha \bm{N_r}^T \bm{\Bar{q}_r}.
\end{equation}
By rearranging terms, Equation (\ref{eq:modified_traction}) can be written as:
\begin{align} \label{eq:neumann_imposition}
  \nonumber  \bm{q}^1 & = \{\bm{I} + \alpha \bm{N_r}^T \bm{N_r}\}^{-1} \{\bm{K_a}^1 \bm{u} + \alpha \bm{N_r}^T \bm{\Bar{q}_r} \} \\ 
        & = \bm{Q}^{-1} \{\bm{K}_a^1 \bm{u} + \alpha \bm{N_r}^T \bm{\Bar{q}_r} \},
\end{align}
where $\bm{I}$ is the identity matrix and $\bm{Q} = \bm{I} + \alpha \bm{N_r}^T \bm{N_r}$. 

Combining Equations (\ref{eq:heat_balance_nodal_mat}) and (\ref{eq:neumann_imposition}), the matrix form of the modified heat transfer balance equation is given by:
\begin{equation} \label{eq:heat_balance_nodal_mat_mod}
    \bm{M} \dot{\bm{u}} + \bm{K' u} = \bm{f} - \alpha \bm{K_s^1} \bm{Q}^{-1} \bm{N_r}^T \bm{\Bar{q}_r},
\end{equation}
where $\quad \bm{K'} = \bm{K_s^1 Q^{-1} K_a^1} + \bm{K_s^2 K_a^2}$. In the penalty method a large value should be selected for the penalty factor $\alpha$ to ensure the accuracy of the BC enforcement. However, if $\alpha$ is too large, stability issues may arise. In our study, we found that $\alpha$ in the range  $\left[10^4, 10^7\right]$ led to satisfactory results, with the best ones (lowest approximation errors in benchmark problems) obtained for $\alpha = 10^6$.

\subsection{Computer implementation} \label{subsec:rpimc_implementation}
Regularly distributed nodes with equidistant spacing $h$ in all coordinates were considered. The RPI shape parameters were selected as $\alpha_c = 1.5$, $d_c = h$ and $q = 1.03$. The penalty factor $\alpha = 10^6$ was chosen to enforce Neumann BCs through the penalty method. The standard forward finite difference scheme (forward Euler) with mass lumping was used to approximate partial differentiation with respect to time explicitly. The forward Euler method is well-known as a conditionally stable method. To ensure stability, an adequately small time step must be used. An estimation of the stable time step was computed by applying the Gerschg\"{o}rin theorem \cite{Myers1978}:
\begin{equation} \label{eq:gerschgorin}
    dt_s = \min_{i=1,...,n} \left[ \frac{m_{ii}}{k_{ii} + \sum\limits_{\substack{j=1 \\ j \neq i}}^{n} \abs{k_{ij}}} \right].
\end{equation}
where $m_{ii}$, $k_{ii}$ are the diagonal entries in $\bm{M}$ and $\bm{K'}$ matrices, respectively. The selected time step $dt = (0.9) dt_s$ was chosen after applying a $10\%$ reduction to the stable time step to ensure the stability of the time integration. The pseudo-code of the RPIMC method's computer implementation is given in Algorithm (\ref{alg:implementation}).

\begin{algorithm}[H]
  \caption{Radial Point Interpolation Mixed Collocation (RPIMC) algorithm} \label{alg:implementation}
  \begin{algorithmic}[1]
    \Procedure{RPIMC}{$\Omega, t_f$} \Comment{The RPIMC solution in domain $\Omega$ for time [0, $t_f$]}
      \State \texttt{initialize field variable: $\bm{u} = \bm{0}$}
      \State \texttt{distribute field nodes in domain $\Omega$}
      \State \texttt{compute normals for boundary field nodes}
      \For{\texttt{<each field node $i$>}}
        \State \texttt{find field nodes in the local support domain of $i$}
        \State \texttt{compute basis functions and derivatives}
        \State \texttt{assemble matrices: $\bm{K_s^1}$, $\bm{K_s^2}$, $\bm{K_a^1}$, $\bm{K_a^2}$, $\bm{M}$}
        \If {\texttt{<$i$ is on Neumann boundary>}}
            \State \texttt{assemble matrices: $\bm{N_r}$, $\bm{Q}$}
        \EndIf
      \EndFor
      \State \texttt{assemble matrix: $\bm{K'}$}
      \State \texttt{compute $dt$} \Comment{Using Gerschg\"{o}rin Theorem, Equation (\ref{eq:gerschgorin})}
      \While {\texttt{<$t <= t_{f}$>}}
        \State \texttt{update body source: $\bm{f}$}
        \State \texttt{update field variable: $\bm{u}$} \Comment{Using Forward Euler scheme}
        \State \texttt{$t = t + dt$}
      \EndWhile
    \EndProcedure
  \end{algorithmic}
\end{algorithm}

\section{Numerical Benchmarks and Cardiac Electrophysiology problem} \label{sec:benchmark}
The performance of the RPIMC method is presented for several 2D and 3D heat transfer benchmark problems for which an analytical solution is available. Convergence analysis for the numerical solution $u^h$ against the analytical solution $u^{an}$ was performed in terms of the $E_2$ and $NRMS$ error metrics given by:
\begin{equation} \label{eq:error_metrics}
\begin{aligned}
    E_2 &= \left( \frac{\sum_{\bm{x}_i \in \Omega} (u^h(\bm{x}_i) - u^{an}(\bm{x_i}))^2}{\sum_{\bm{x}_i \in \Omega} u^{an}(\bm{x}_i)^2}  \right)^{1/2}, \\
    NRMS &= \frac{\left( \sum_{\bm{x}_i \in \Omega} \left(u^h(\bm{x}_i) - u^{an}(\bm{x}_i) \right)^2   \right)^{1/2}}{max \left|u^{an}(\bm{x}_i) \right| - min \left|u^{an}(\bm{x}_i) \right|}.
\end{aligned}
\end{equation}

For comparison, the benchmark problems were additionally solved with the MLPG-MC and FEM methods and convergence analysis was performed. The convergence rate ($\Bar{\rho}$) for the $E_2$ and $NRMS$ error metrics at successive refinements was calculated at the final simulation time $t = t_{f}$ using Equation (\ref{eq:crate}), as proposed in \cite{Thamareerat2016}:
\begin{equation} \label{eq:crate}
    \Bar{\rho} = \frac{\log(\frac{E_a}{E_b})}{\log(\frac{h_a}{h_b})}
\end{equation}
\noindent where $E_a$, $E_b$ denote the error and $h_a$, $h_b$ the nodal spacing at two successive refinements. For the MLPG-MC method, the MLS basis function with linear polynomial basis was used as trial function and the quartic spline function as test function. FEM simulations were performed by using linear triangle and tetrahedral elements in 2D and 3D problems, respectively.

The RPI and MLS approximation schemes used in this work were implemented using MATLAB and are available in an open-source repository \cite{mountris2020mfree}.

\subsection{Lateral heat loss in 2D with Dirichlet boundary conditions} \label{subsec:heat2d_dirichlet}
A heat conduction problem with lateral heat loss was solved in a 2D square domain $\Omega$ with edge length $l=1$. The problem is described by the PDE:
\begin{equation} \label{eq:heat_2d_dirichlet_pde}
    u_{,0} = u_{,xx} + u_{,yy} + (1+t^2)u + (2 \pi^2 - t^2 - 2) \cross sin(\pi x)  cos(\pi y), \quad (x,y) \in \Omega, \quad t > 0
\end{equation}
with Dirichlet BCs on $\partial \Omega$:
\begin{equation} \label{eq:heat_2d_dirichlet_bc}
\begin{aligned}
    u(0,y,t) &= u(1,y,t) = 0, \\
    u(x,0,t) &= u(x,1,t) = e^{-t} \text{sin}(\pi x),
\end{aligned}
\end{equation}
The initial condition was obtained by the analytical solution for $t=0$:
\begin{equation} \label{eq:heat_2d_dirichlet_an}
    u(x,y,t) = e^{-t} \text{sin}(\pi x) \text{cos}(\pi y).
\end{equation}

\begin{figure}[H]
    \centering
    \begin{minipage}{0.45\linewidth}
        \centering
        \includegraphics[width=\textwidth]{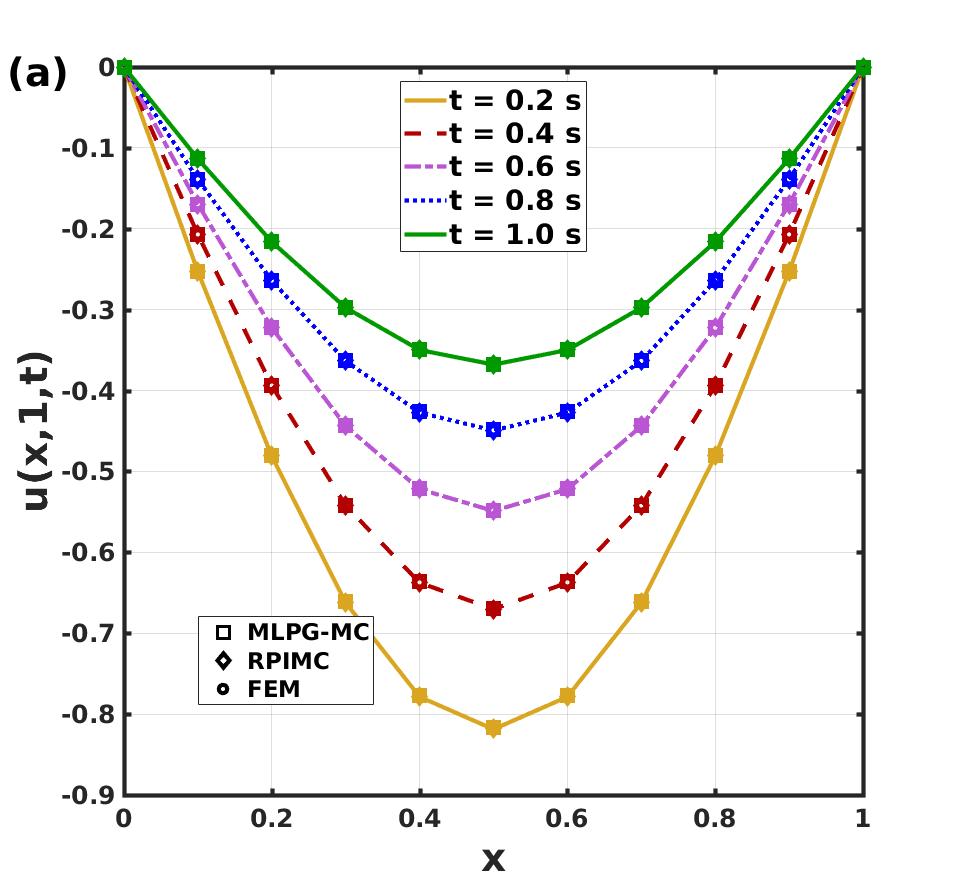}
    \end{minipage}
    \hspace{0.5cm}
    \begin{minipage}{0.45\linewidth}
        \centering
        \includegraphics[width=\textwidth]{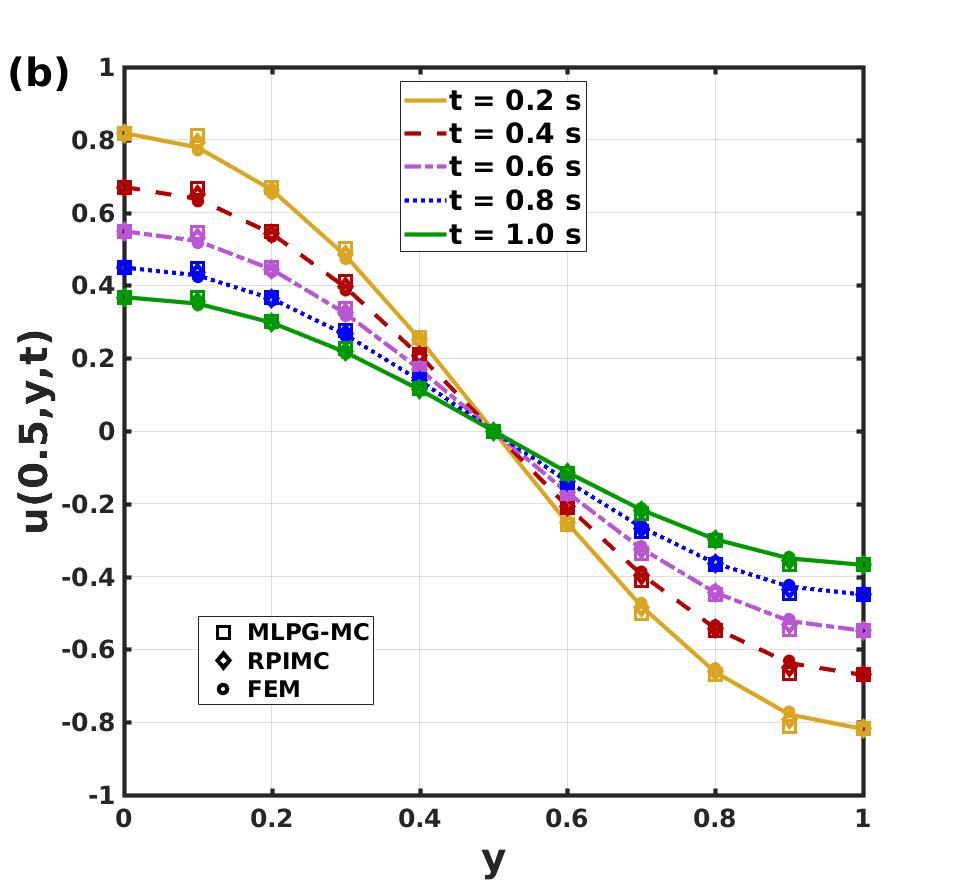}
    \end{minipage}
    \caption{Solution for heat conduction with lateral heat loss in 2D for $t = [0, 1]$ using RPIMC ($\Diamond$), MLPG-MC ($\square$), FEM ($\bigcirc$). Plotted lines correspond to the analytical solution. (a) Solution profile for $y = 1$. (b) Solution profile for $x = 0.5$.}
    \label{fig:heat_2d_profiles}
\end{figure}

\begin{figure}[H]
    \centering
    \begin{minipage}{0.45\linewidth}
        \centering
        \includegraphics[width=\textwidth]{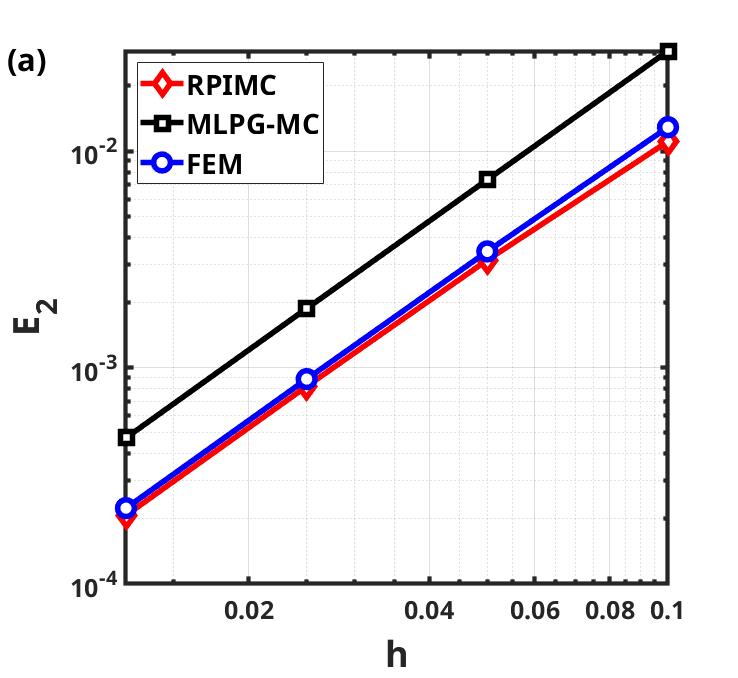}
    \end{minipage}
    \hspace{0.5cm}
    \begin{minipage}{0.45\linewidth}
        \centering
        \includegraphics[width=\textwidth]{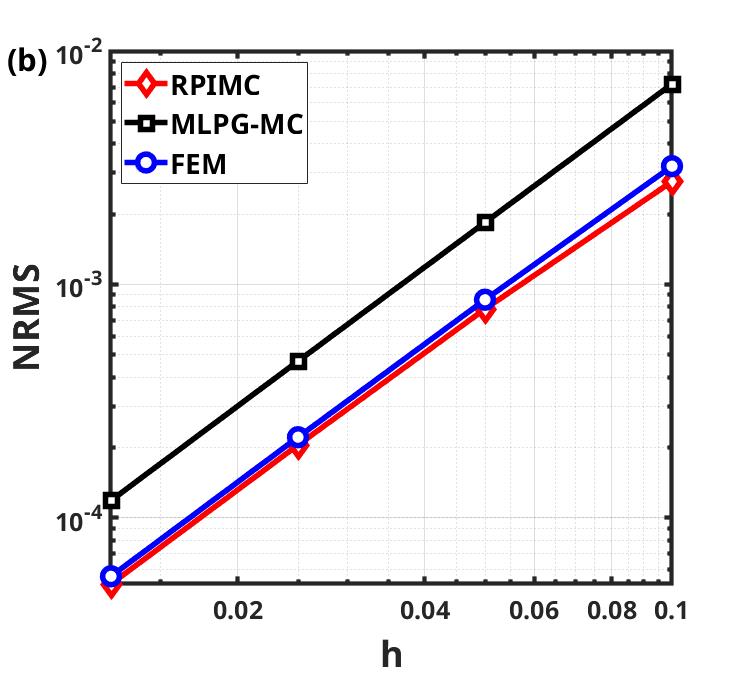}
    \end{minipage}
    \caption{Convergence for heat conduction with lateral heat loss in 2D at $t = 1$ using RPIMC ($\Diamond$), MLPG-MC ($\square$), FEM ($\bigcirc$). (a) Convergence in $E_2$ error metric. (b) Convergence in $NRMS$ error metric.}
    \label{fig:heat_2d_convergence}
\end{figure}

The problem was solved for the time interval $t = [0, 1]$ for $11 \cross 11$ regularly distributed nodes in $\Omega$ with spatial spacing $h=0.1$. Figure (\ref{fig:heat_2d_profiles}) shows the profiles of the solution for $y = 1$ and for $x = 0.5$, respectively. Convergence analysis was performed for successive refinements with $h = [0.1, 0.05, 0.025, 0.0125]$. The convergence analysis results for the $E_2$ and $NRMS$ error metrics are presented in Figure (\ref{fig:heat_2d_convergence}). A summary of the convergence rates for $E_2$ and $NRMS$ error metrics is provided in Table \ref{tab:e2_summary} and Table \ref{tab:nrms_summary}.

\subsection{Heat conduction in 3D with insulated borders} \label{subsec:heat3d_neumann}
A heat conduction problem with insulated borders was solved in a 3D cubic domain $\Omega$ with edge length $l = \pi$. The problem is governed by the PDE:
\begin{equation} \label{eq:heat_3d_neumann_pde}
    u_{,0} = u_{,xx} + u_{,yy} + u_{,zz}; \quad (x,y,z) \in \Omega, \quad t > 0,
\end{equation}
\noindent with Neumann BCs on $\partial \Omega$:
\begin{equation} \label{eq:heat_3d_neumann_bc}
    u_{,x} \vert _{x=0} = u_{,x} \vert _{x=\pi} = u_{,y} \vert _{y=0} = u_{,y} \vert _{y=\pi} = u_{,z} \vert _{z=0} = u_{,z} \vert _{z=\pi} = 0,
\end{equation}
The initial condition was obtained by the analytical solution for $t=0$:
\begin{equation}
    u(x,y,z,t) = 1 + 2 e^{-3t} \text{cos}(x) \text{cos}(y) \text{cos}(z) + 3 e^{-29t} \text{cos}(2x) \text{cos}(3y) \text{cos}(4z).
\end{equation}

\begin{figure}[H]
    \centering
    \begin{minipage}{0.45\linewidth}
        \centering
        \includegraphics[width=\textwidth]{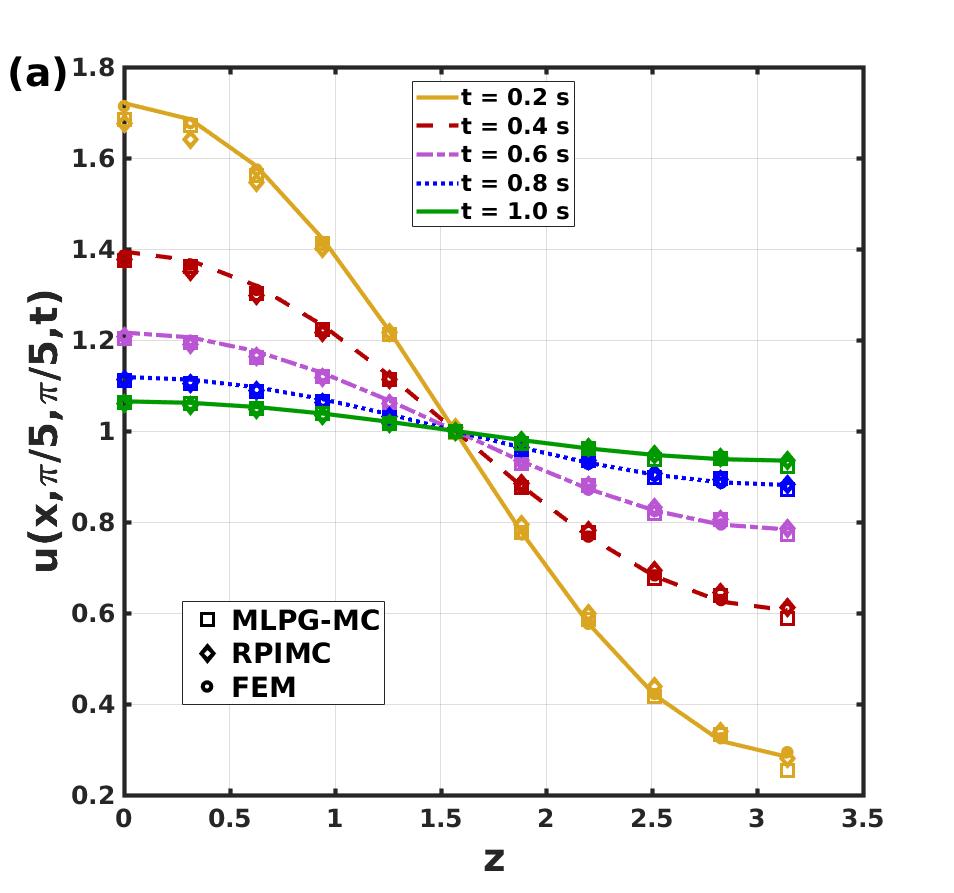}
    \end{minipage}
    \hspace{0.5cm}
    \begin{minipage}{0.45\linewidth}
        \centering
        \includegraphics[width=\textwidth]{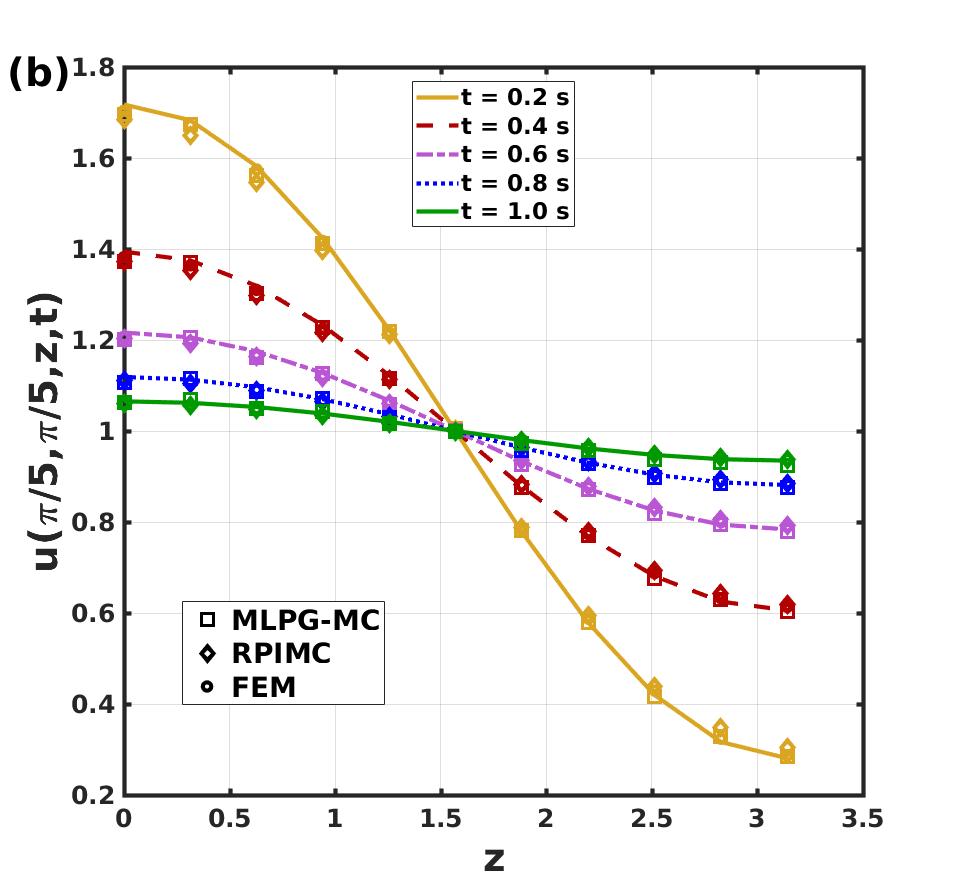}
    \end{minipage}
    \caption{Solution for heat conduction with insulated borders in 3D for $t = [0, 1]$ using RPIMC ($\Diamond$), MLPG-MC ($\square$), FEM ($\bigcirc$). Plotted lines correspond to the analytical solution. (a) Solution profile for $y = z = \pi/5$. (b) Solution profile for $x = y = \pi/5$.}
    \label{fig:heat_3d_neumann}
\end{figure}

\begin{figure}[H]
    \centering
    \begin{minipage}{0.45\linewidth}
        \centering
        \includegraphics[width=\textwidth]{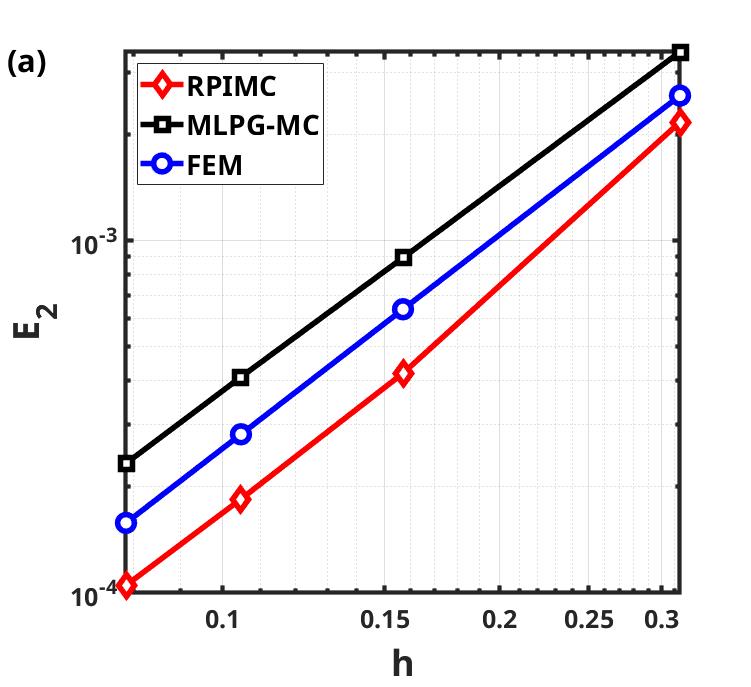}
    \end{minipage}
    \hspace{0.5cm}
    \begin{minipage}{0.45\linewidth}
        \centering
        \includegraphics[width=\textwidth]{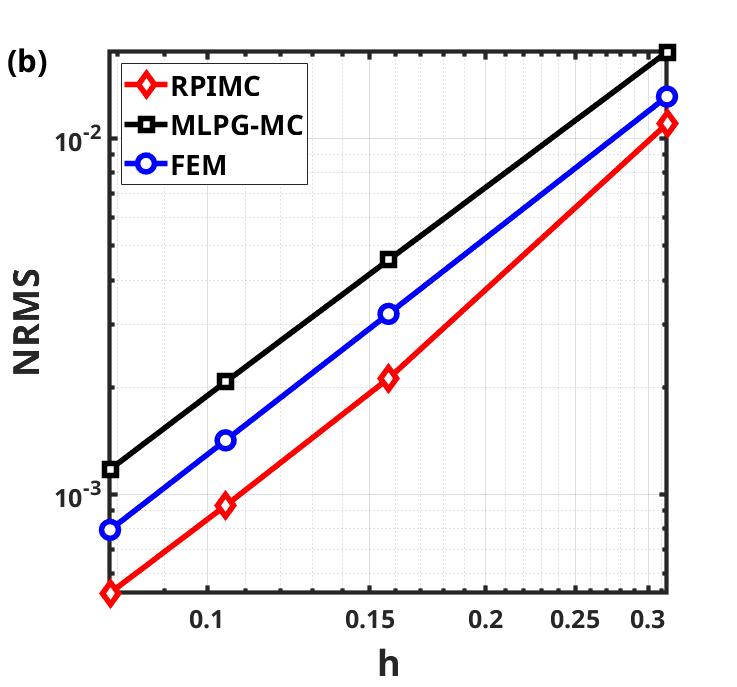}
    \end{minipage}
    \caption{Convergence for heat conduction with insulated borders in 3D at $t = 1$ using RPIMC ($\Diamond$), MLPG-MC ($\square$), FEM ($\bigcirc$). (a) Convergence in $E_2$ error metric. (b) Convergence in $NRMS$ error metric.}
    \label{fig:heat_insulated3d_convergence}
\end{figure}

The problem was solved for the time interval $t = [0,1]$ for $11 \cross 11$ regularly distributed nodes in $\Omega$ with spatial spacing $h = \pi/10$. Figure (\ref{fig:heat_3d_neumann}) shows the profiles of the solution for $y = z = \pi/5$ and $x = y = \pi/5$. Convergence analysis was performed for successive refinements with $h = [\pi/10,\pi/20,\pi/30,\pi/40]$. The convergence analysis results for the $E_2$ and $NRMS$ error metrics are presented in Figure(\ref{fig:heat_insulated3d_convergence}). A summary of the convergence rates for $E_2$ and $NRMS$ error metrics is provided in Table \ref{tab:e2_summary} and Table \ref{tab:nrms_summary}.

\subsection{Inhomogeneous heat conduction in 3D with Dirichlet boundary conditions} \label{subsec:heat3d_inhom}
An inhomogeneous heat conduction problem with Dirichlet BCs was solved in a 3D cubic domain with edge length $l=\pi$. The problem is described by the following PDE:
\begin{equation} \label{eq:heat_3d_inhom_pde}
    u_{,0} = u_{,xx} + u_{,yy} + u_{,zz} + \text{sin} (z); \quad 0 < x,y,z < \pi, \quad t > 0,
\end{equation}
with the following Dirichlet BCs:
\begin{align} \label{eq:heat_3d_inhom_bc}
    u(0,y,z,t) &= \text{sin} (z) + e^{-2t} \text{sin} (y) \\
    u(\pi,y,z,t) &= \text{sin} (z) - e^{-2t} \text{sin} (y) \\
    u(x,0,z,t) &= \text{sin} (z) + e^{-2t} \text{sin} (x) \\
    u(x,\pi,z,t) &= \text{sin} (z) - e^{-2t} \text{sin} (x) \\
    u(x,y,0,t) &= u(x,y,\pi,t) = e^{-2t} \text{sin} (x + y)
\end{align}
The initial condition was obtained by the analytical solution for $t=0$:
\begin{equation} \label{eq:heat_3d_inhom_init}
    u(x,y,z,0) = \text{sin} (z) + e^{-2t} \text{sin} (x+y).
\end{equation}

\begin{figure}[H]
    \centering
    \begin{minipage}{0.45\linewidth}
        \centering
        \includegraphics[width=\textwidth]{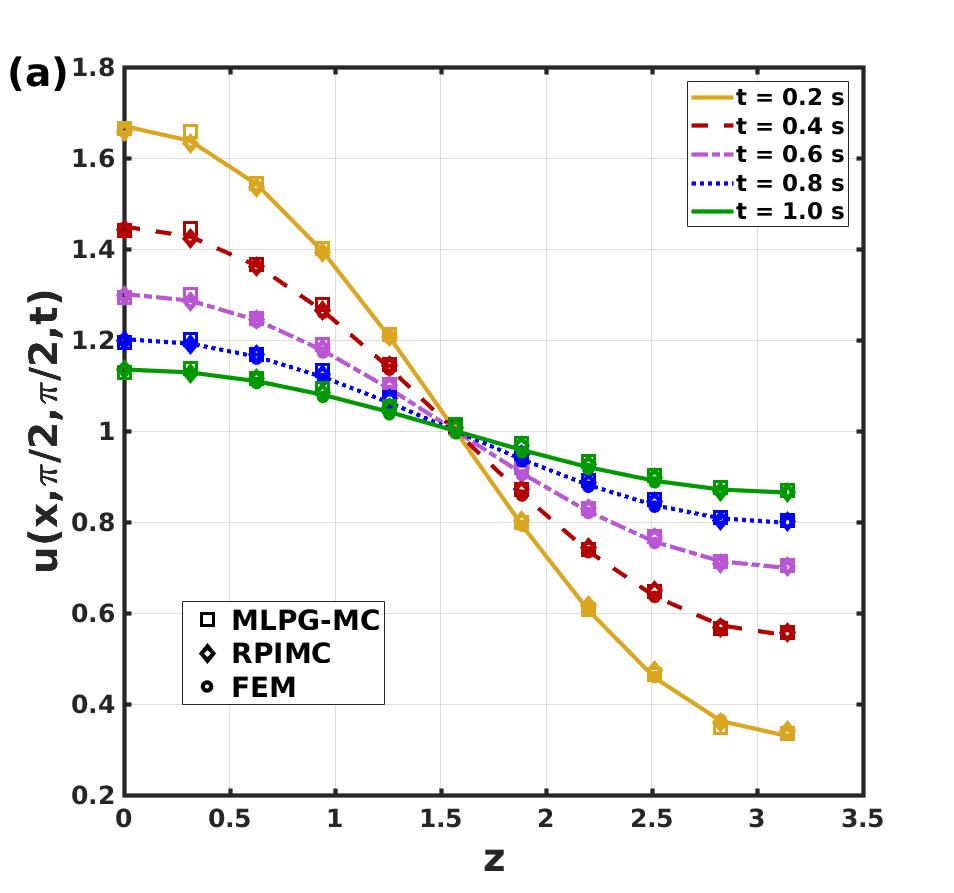}
    \end{minipage}
    \hspace{0.5cm}
    \begin{minipage}{0.45\linewidth}
        \centering
        \includegraphics[width=\textwidth]{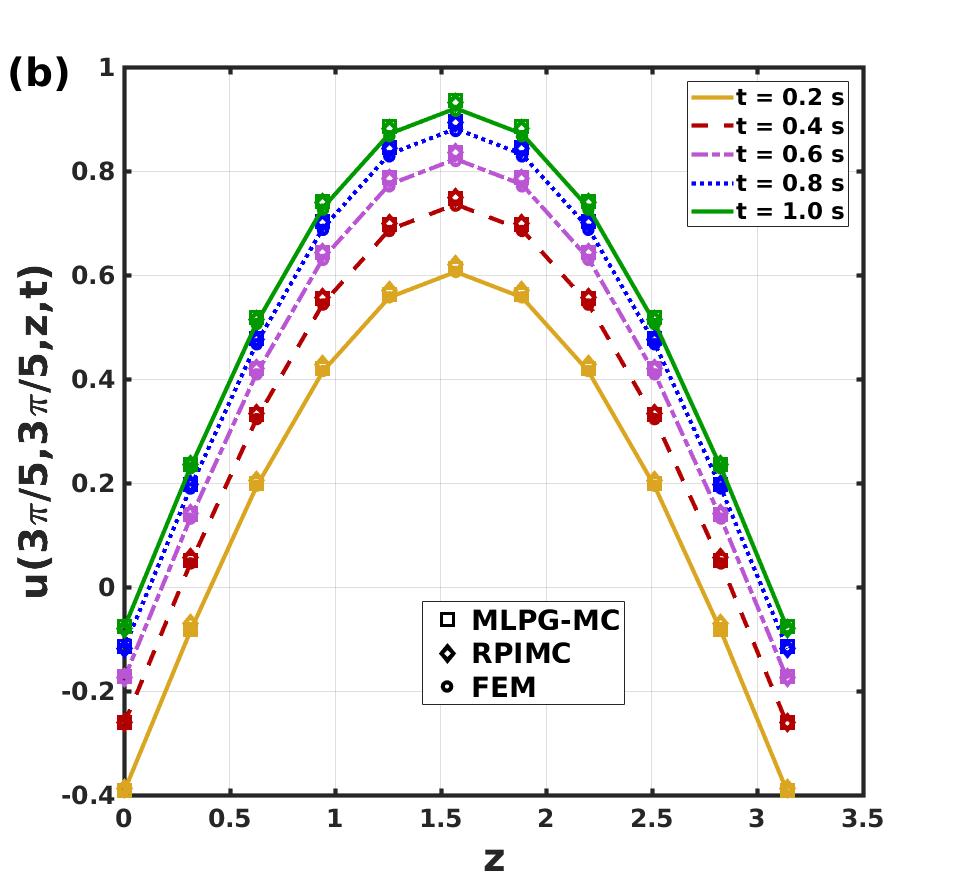}
    \end{minipage}
    \caption{Solution for inhomogeneous heat conduction in 3D with Dirichlet boundary conditions for $t = [0, 1]$ RPIMC ($\Diamond$), MLPG-MC ($\square$), FEM ($\bigcirc$). Plotted lines correspond to the analytical solution. (a) Solution profile for $y = z = \pi/2$. (b) Solution profile for $x = y = 3\pi/5$.}
  \label{fig:heat_3d_inhom}
\end{figure}

\begin{figure}[H]
    \centering
    \begin{minipage}{0.45\linewidth}
        \centering
        \includegraphics[width=\textwidth]{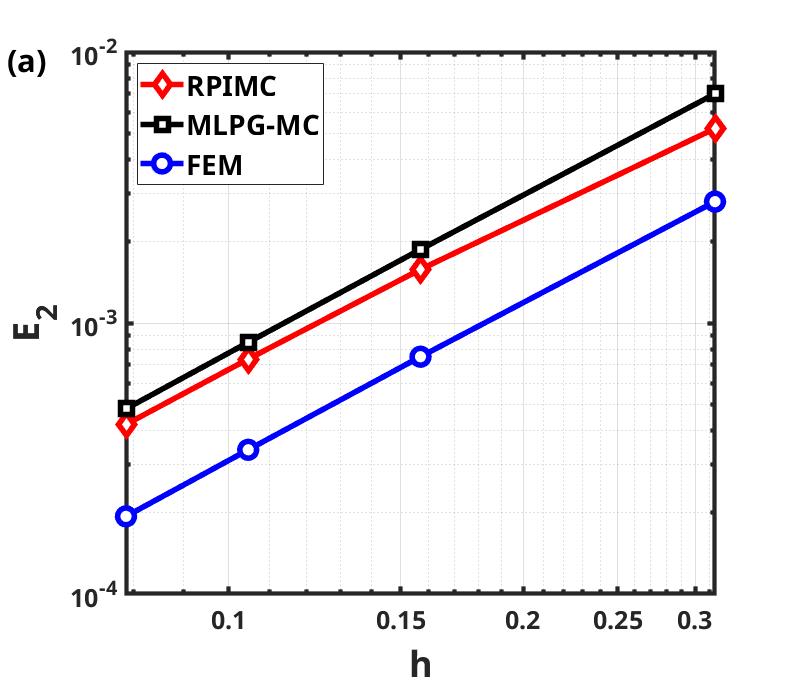}
    \end{minipage}
    \hspace{0.5cm}
    \begin{minipage}{0.45\linewidth}
        \centering
        \includegraphics[width=\textwidth]{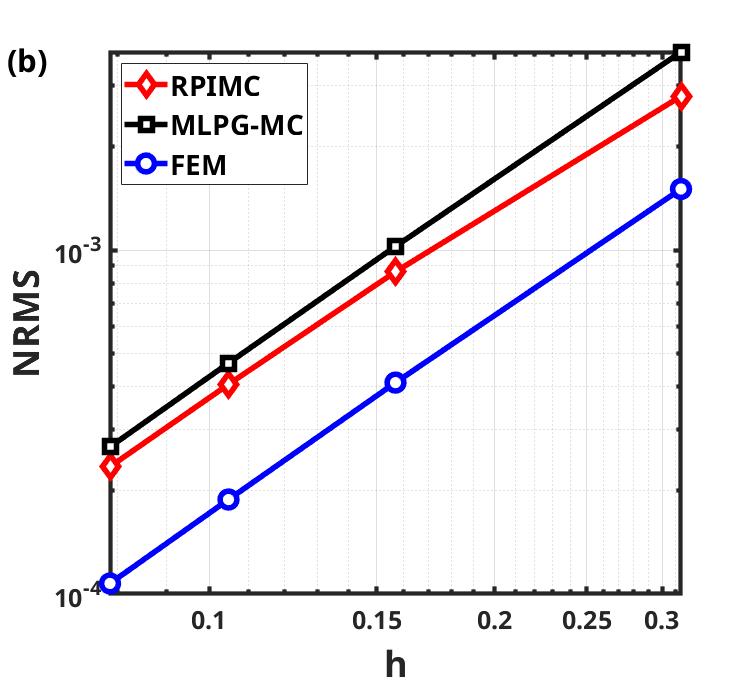}
    \end{minipage}
    \caption{Convergence for inhomogeneous heat conduction in 3D with Dirichlet boundary conditions at $t = 1$ using RPIMC ($\Diamond$), MLPG-MC ($\square$), FEM ($\bigcirc$). (a) Convergence in $E_2$ error metric. (b) Convergence in $NRMS$ error metric.}
    \label{fig:heat_inhom3d_convergence}
\end{figure}

The problem was solved for the time interval $t = [0,1]$ for $11 \cross 11$ regularly distributed nodes in $\Omega$ with spatial spacing $h = \pi/10$. Figure (\ref{fig:heat_3d_inhom}) shows the profiles of the solution for $y = z = \pi/2$ and for $x = y = 3\pi/5$. Convergence analysis was performed for successive refinements with $h = [\pi/10,\pi/20,\pi/30,\pi/40]$. The convergence analysis results for the $E_2$ and $NRMS$ error metrics are presented in Figure(\ref{fig:heat_inhom3d_convergence}). A summary of the convergence rates for $E_2$ and $NRMS$ error metrics is provided in Table \ref{tab:e2_summary} and Table \ref{tab:nrms_summary}.

\begin{table}[H]
\footnotesize
\centering
\caption{\label{tab:e2_summary} Summary of $E_2$ error metrics and convergence rates for all benchmark problems.}
\begin{tabular}{ l c c c c c c } 
    \hline
    h & \multicolumn{3}{c}{$E_2$ error} & \multicolumn{3}{c}{Convergence rate ($\Bar{\rho}$)} \\
      &  RPIMC & MLPG-MC & FEM & RPIMC & MLPG-MC & FEM \\ 
    \hline \hline
    \multicolumn{7}{l}{Benchmark \ref{subsec:heat2d_dirichlet}: Lateral heat loss in 2D with Dirichlet boundary conditions} \\ \hline
    0.1    & 1.10 x $10^{-2}$ & 2.90 x $10^{-2}$ & 1.29 x $10^{-2}$ & -    & -    & - \\
    0.05   & 3.14 x $10^{-3}$ & 7.37 x $10^{-3}$ & 3.40 x $10^{-3}$ & 1.82 & 1.96 & 1.91 \\
    0.025  & 8.19 x $10^{-4}$ & 1.87 x $10^{-3}$ & 9.00 x $10^{-4}$ & 1.94 & 1.97 & 1.96 \\
    0.0125 & 2.07 x $10^{-4}$ & 4.73 x $10^{-4}$ & 2.00 x $10^{-4}$ & 1.98 & 1.98 & 1.98 \\
    \hline \hline
    \multicolumn{7}{l}{Benchmark \ref{subsec:heat3d_neumann}: Heat conduction in 3D with insulated borders} \\ \hline
    0.314  & 2.17 x $10^{-3}$ & 3.43 x $10^{-3}$ & 2.60 x $10^{-3}$ & -    & -    & - \\
    0.157  & 4.21 x $10^{-4}$ & 8.95 x $10^{-4}$ & 6.00 x $10^{-4}$ & 2.37 & 1.94 & 2.02 \\
    0.108  & 1.85 x $10^{-4}$ & 4.10 x $10^{-4}$ & 3.00 x $10^{-4}$ & 2.03 & 1.93 & 2.02 \\
    0.079  & 1.05 x $10^{-4}$ & 2.33 x $10^{-4}$ & 2.00 x $10^{-4}$ & 1.97 & 1.96 & 2.01 \\
    \hline \hline
    \multicolumn{7}{l}{Benchmark \ref{subsec:heat3d_inhom}: Inhomogeneous heat conduction in 3D with Dirichlet boundary conditions} \\ \hline
    0.314  & 5.20 x $10^{-3}$ & 7.00 x $10^{-3}$ & 2.80 x $10^{-3}$ & -    & -    & - \\
    0.157  & 1.60 x $10^{-3}$ & 1.90 x $10^{-3}$ & 8.00 x $10^{-4}$ & 1.73 & 1.91 & 1.90 \\
    0.108  & 7.40 x $10^{-4}$ & 8.50 x $10^{-4}$ & 3.00 x $10^{-4}$ & 1.88 & 1.95 & 1.95 \\
    0.079  & 4.20 x $10^{-4}$ & 4.80 x $10^{-4}$ & 2.00 x $10^{-4}$ & 1.93 & 1.96 & 1.97 \\
\hline
\end{tabular}
\end{table}

\begin{table}[H]
\footnotesize
\centering
\caption{\label{tab:nrms_summary} Summary of NRMS error metrics and convergence rates for all benchmark problems.}
\begin{tabular}{ l c c c c c c } 
    \hline
    h & \multicolumn{3}{c}{NRMS error} & \multicolumn{3}{c}{Convergence rate ($\Bar{\rho}$)} \\
      & RPIMC & MLPG-MC & FEM & RPIMC & MLPG-MC & FEM \\ 
    \hline \hline
    \multicolumn{7}{l}{Benchmark \ref{subsec:heat2d_dirichlet}: Lateral heat loss in 2D with Dirichlet boundary conditions} \\ \hline
    0.1    & 2.76 x $10^{-3}$ & 7.22 x $10^{-3}$ & 3.20 x $10^{-3}$ & -    & -    & - \\
    0.05   & 7.83 x $10^{-4}$ & 1.84 x $10^{-3}$ & 8.58 x $10^{-4}$ & 1.82 & 1.94 & 1.90 \\
    0.025  & 2.04 x $10^{-4}$ & 4.68 x $10^{-4}$ & 2.21 x $10^{-4}$ & 1.94 & 1.97 & 1.96 \\
    0.0125 & 5.19 x $10^{-5}$ & 1.18 x $10^{-4}$ & 5.60 x $10^{-5}$ & 1.98 & 1.98 & 1.98 \\
    \hline \hline
    \multicolumn{7}{l}{Benchmark \ref{subsec:heat3d_neumann}: Heat conduction in 3D with insulated borders} \\ \hline
    0.314  & 1.10 x $10^{-2}$ & 1.75 x $10^{-2}$ & 1.31 x $10^{-2}$ & -    & -    & - \\
    0.157  & 2.11 x $10^{-3}$ & 4.55 x $10^{-3}$ & 3.21 x $10^{-3}$ & 2.37 & 1.94 & 2.03 \\
    0.108  & 9.30 x $10^{-4}$ & 2.07 x $10^{-3}$ & 1.42 x $10^{-3}$ & 2.03 & 1.94 & 2.02 \\
    0.079  & 5.28 x $10^{-4}$ & 1.17 x $10^{-3}$ & 7.95 x $10^{-4}$ & 1.96 & 1.97 & 2.01 \\
    \hline \hline
    \multicolumn{7}{l}{Benchmark \ref{subsec:heat3d_inhom}: Inhomogeneous heat conduction in 3D with Dirichlet boundary conditions} \\ \hline
    0.314  & 2.80 x $10^{-3}$ & 3.76 x $10^{-3}$ & 1.50 x $10^{-3}$ & -    & -    & - \\
    0.157  & 8.65 x $10^{-4}$ & 1.02 x $10^{-4}$ & 4.12 x $10^{-4}$ & 1.70 & 1.88 & 1.87 \\
    0.108  & 4.10 x $10^{-4}$ & 4.69 x $10^{-4}$ & 1.88 x $10^{-4}$ & 1.86 & 1.93 & 1.93 \\
    0.079  & 2.30 x $10^{-4}$ & 2.68 x $10^{-4}$ & 21.07 x $10^{-4}$ & 1.92 & 1.95 & 1.95 \\
\hline
\end{tabular}
\end{table}

\subsection{Electrical propagation in a cardiac biventricular model} \label{subsec:cardiac_problem}

The propagation of an electrical stimulus in a cardiac biventricular geometry was simulated by solving the decoupled monodomain model after application of the operator splitting method \cite{qu1999advanced} given by:
\begin{equation} \label{eq:monodomain}
\begin{array}{ll}
    \partial V / \partial t = -I_{ion}(V) / C &\textrm{ in } \Omega \\
    \partial V / \partial t = \bm{\nabla} \cdot (\bm{D} \bm{\nabla} V) &\textrm{ in } \Omega \\
    \bm{n} \cdot (\bm{D}\bm{\nabla}V) = 0 &\textrm{ in } \partial \Omega
\end{array}
\end{equation}
\noindent where $\Omega$ and $\partial \Omega$ denote the domain of interest and its boundary, respectively, and $\bm{n}$ is the outward unit vector normal to the boundary. $\partial V / \partial t$ is the time derivative of the transmembrane voltage, $I_{ion}$ is the total ionic current and $C$ is the cell capacitance per unit surface area. $\bm{D}$ denotes the diffusion tensor calculated as:
\begin{equation} \label{eq:diffusin_tensor}
    \bm{D} = d_0 [(1-\rho) \bm{f} \otimes \bm{f} + \rho \bm{I}], 
\end{equation}
\noindent where $d_0$ denotes the diffusion coefficient along the myocardial fiber direction, $\rho$ is the transverse-to-longitudinal ratio of conductivity, $f$ is the myocardial fiber direction vector, $I$ is the identity matrix and $\otimes$ denotes the tensor product operation.

The biventricular anatomy was discretized in a tetrahedral mesh with 273919 nodes and 1334218 elements. The myocardial fiber direction vectors were computed using a rule-based method \cite{doste2019}. The value for the diffusion coefficient in the fiber direction was set to $d_0 = 0.002$ cm$^2$/ms and the value for the transverse-to-longitudinal ratio of conductivity was set to $\rho = 0.25$. The fast conduction system in the biventricular model was generated using a fractal-tree generation algorithm \cite{costabal2016}. Electrical stimulation was applied at the terminal nodes of the fast conduction system, so called Purkinje-Myocardial Junctions (PMJs). Stimuli of 1-ms duration and twice the diastolic threshold in amplitude were applied onto the PMJs at a cycle length of 1 s. A full cycle (t = 1 s) was simulated using the RPIMC and MLPG-MC methods to solve the diffusion term of Equation (\ref{eq:monodomain}) according to Equation (\ref{eq:heat_balance_nodal_T}) with zero Neumann BC. The reaction term was defined by using the O'Hara cell model \cite{ohara2011} to represent human ventricular cellular electrophysiology.

\begin{figure}[H]
    \centering
    \includegraphics[width=\textwidth]{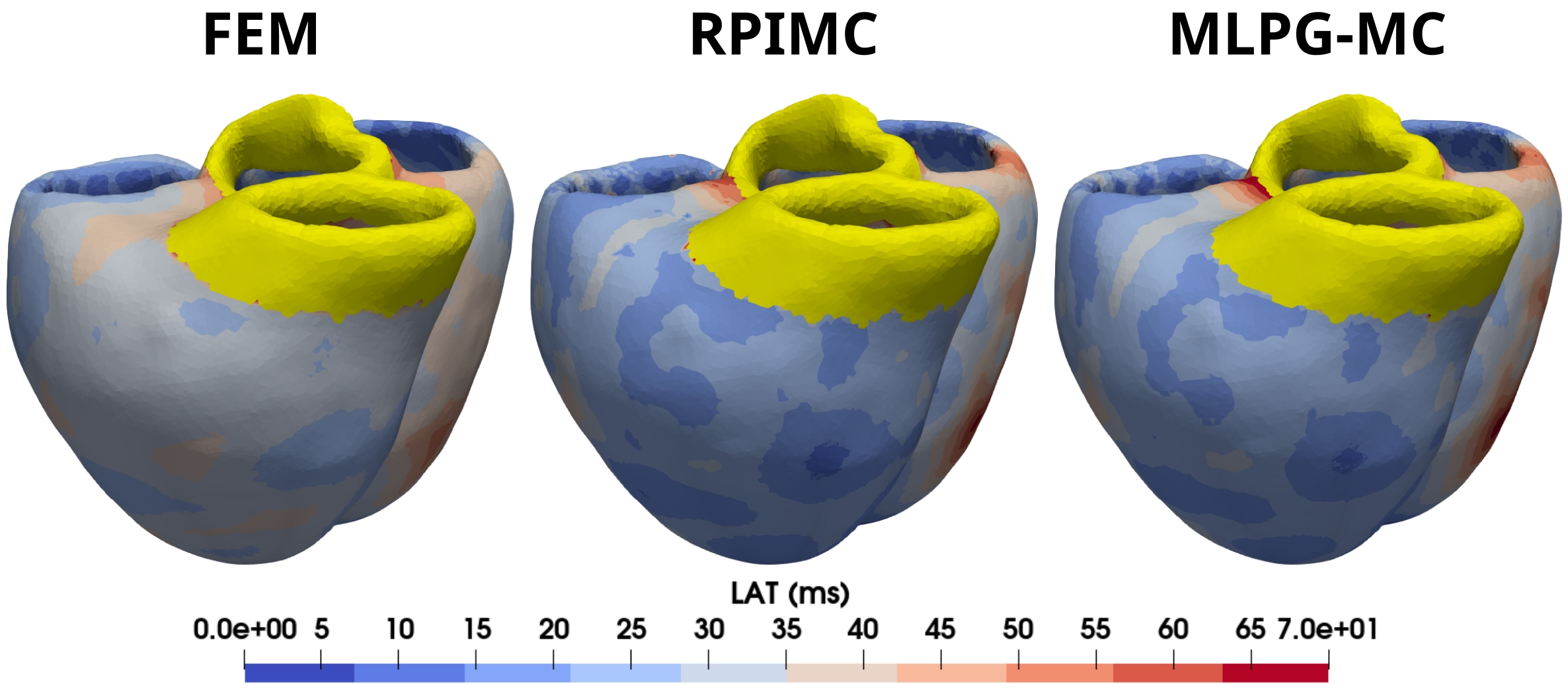}
    \caption{Local activation time maps for a FEM (left), RPIMC (center), and MLPG-MC (right) simulation on a biventricular model. Non-conductive connective tissue is represented in yellow.}
    \label{fig:biventricular_lat}
\end{figure}

A dilatation coefficient set to $a_c = 2.85$ was used for the construction of support domains for both RPIMC and MLPG-MC, leading to support domains containing 51-149 nodes. The RPIMC and MLPG-MC simulation results were compared in terms of local activation time (LAT) with a FEM simulation (Figure \ref{fig:biventricular_lat}). Time integration was performed with the forward Euler method. The critical diffusion step was found to be $dt_{RPIMC} = 0.064$ ms, $dt_{MLPG-MC} = 0.065$ ms, $dt_{FEM} = 0.035$ ms using the Gerschg\"{o}rin theorem \cite{Myers1978}.

Simulations were performed on a laptop with Intel\textsuperscript{\textregistered} Core\texttrademark i7-4720HQ CPU and 16 GB of RAM. The execution time for a full cycle was 99.6 mins for RPIMC, 100.8 mins for MLPG-MC, and 60.6 mins for FEM. From the total execution time, 6.5 mins and 4.4 mins were required for the calculation of the basis functions and their gradients in RPIMC and MLPG-MC, respectively. The mean LAT was found to be 22.4 ms for FEM, 23.3 ms for RPIMC, and 24.0 ms for MLPG-MC. The mean relative difference for the interior nodes LAT was 3.9\% for RPIMC and 7.6\% for MLPG-MC compared to FEM. For the nodes located at the boundary the mean relative difference was 20.4 \% for RPIMC and 22.2 \% for MLPG-MC.

\subsection{Computational efficiency in electrical propagation simulation} \label{subsec:time_profiling}

We evaluated the computational efficiency of the RPIMC method, compared to FEM and MLPG-MC, in a simulation of electrical propagation in a 5 $cm^3$ slab of cardiac tissue composed of epicardial cells. The O'Hara cell model was used to represent human ventricular cellular electrophysiology, as in section \ref{subsec:cardiac_problem}. Myocardial fiber direction was set parallel to the Z-$axis$. The diffusion coefficient along the fibers was set to $d_0= 0.001 cm^2/ms$, with a transverse-to-longitudinal ratio of $\rho= 0.25$. Electrical propagation was simulated for $t = 100$ ms by application of a stimulus at the bottom of the tissue slab (Z = 0 cm) with amplitude twice the diastolic threshold and duration of 1 ms.

Simulations were performed by using FEM, RPIMC and MLPG-MC for four different resolution levels, corresponding to nodal spacing of $h = (0.5, 0.25, 0.125 , 0.075)$ mm, on a laptop with Intel\textsuperscript{\textregistered} Core\texttrademark i7-4720HQ CPU and 16 GB of RAM. The construction of the support domain in RPIMC and MPLPG-MC was performed by using values for the dilatation coefficient of $a_c=2.1$ and $a_c=2.5$ to assess the effect of the support domain size.

\begin{figure}[H]
    \centering
    \begin{minipage}{0.45\linewidth}
        \centering
        \includegraphics[width=\textwidth]{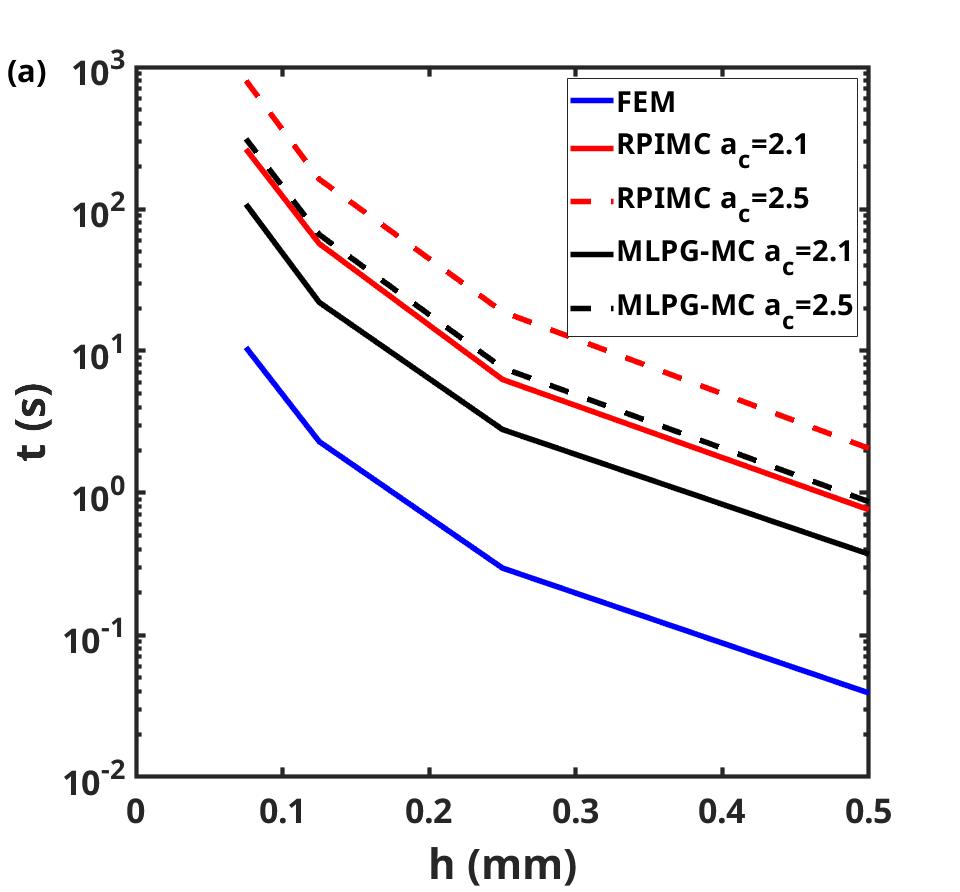}
    \end{minipage}
    \hspace{0.5cm}
    \begin{minipage}{0.45\linewidth}
        \centering
        \includegraphics[width=\textwidth]{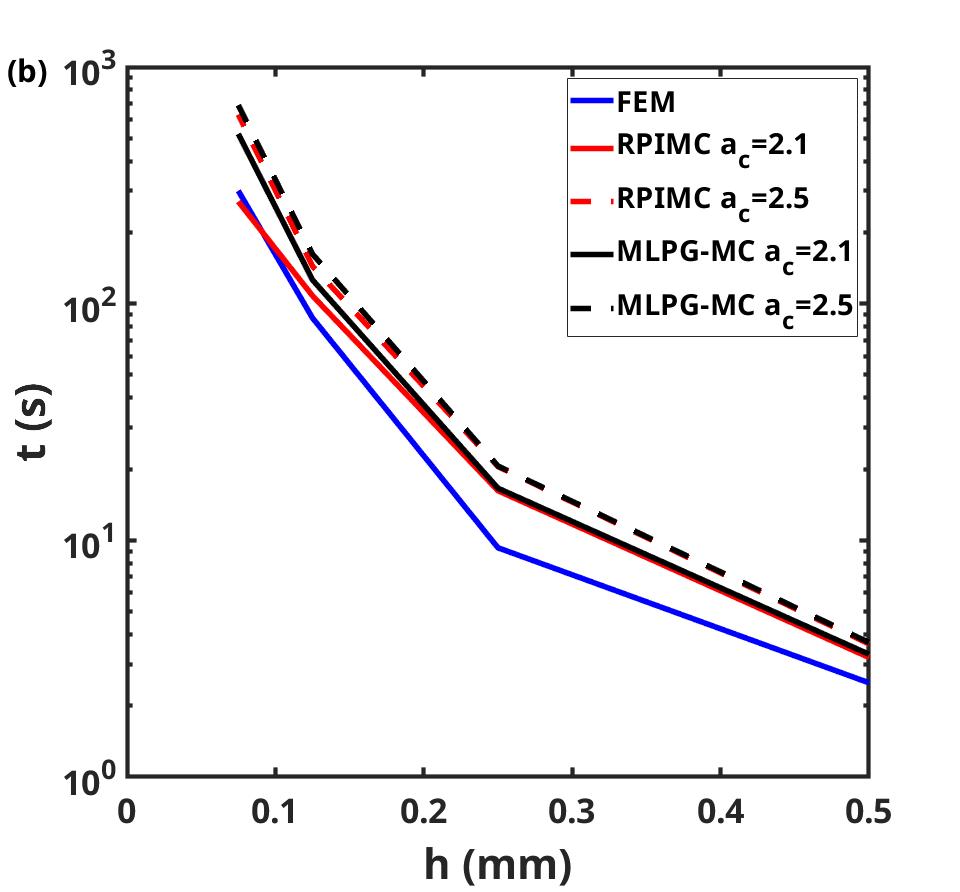}
    \end{minipage}
    \caption{Time profiling for simulation of electrical propagation in a 5 $cm^3$ slab of cardiac epicardial tissue using FEM (blue), RPIMC with $a_c=2.1$ (red), RPIMC with $a_c=2.5$ (red dashed), MLPG-MC with $a_c=2.1$ (black) and MLPG-MC with $a_c=2.5$ (black dashed). (a) Time for stiffness matrix assembly versus nodal spacing $h$. (b) Time for solving the monodomain model versus nodal spacing $h$.}
    \label{fig:time_efficiency}
\end{figure}

The maximum time, corresonding to h=0.075 mm, for stiffness matrix assembly was 11 s for FEM, 264 s for RPIMC with $a_c=2.1$, 804 s for RPIMC with $a_c=2.5$, 108 s for MLPG-MC with $a_c=2.1$ and 312 s for MLPG-MC with $a_c=2.5$. The maximum time for solving the monodomain model was 300 s for FEM, 270 s for RPIMC with $a_c=2.1$, 630 s for RPIMC with $a_c=2.5$, 522 s for MLPG-MC with $a_c=2.1$ and 690 s for MLPG-MC with $a_c=2.5$. The time required to assemble the stiffness matrix and solve the monodomain model for each resolution level is reported in Figure \ref{fig:time_efficiency}. RPIMC required longer time than MLPG-MC for stiffness matrix assembly. This was due to the inversion operation of the enriched moment matrix $\bm{G}$ during the computation of the RPI basis functions and their gradients. However, the time required for the solution of the monodomain model with RPIMC was smaller for all resolution levels.

\section{Concluding remarks} \label{sec:conclusions}

The RPIMC method was proposed and tested to solve transient diffusion problems. Since RPIMC uses RPI basis functions as trial functions that possess the delta Kronecker property, Dirichlet BCs were imposed similarly to FEM. The results obtained for a number of benchmark problems demonstrated that RPIMC can achieve high accuracy, similar to that of FEM, generally outperforming the MLPG-MC method.

Furthermore, we showed that RPIMC can be used to solve the monodomain model for simulation of electrical propagation in the heart. Local activation time maps obtained by RPIMC were found to be in good agreement with those of FEM, being remarkably closer than those obtained by MLPG-MC. However, deterioration of the RPIMC solution at the boundary nodes, where Neumann boundary conditions were imposed, was observed. Rough edges present on the biventricular model's surfaces led to discontinuities in normal vectors' direction, which is postulated to have negatively affected the accuracy at the Neumann boundary.

In terms of computational efficiency, RPIMC and MLPG-MC methods performed similarly in simulations of electrical propagation in a cardiac tissue slab, being somewhat less efficient than FEM. As expected, the time for the assembly of the stiffness matrix was higher for RPIMC and MLPG-MC than for FEM due to the additional workload associated with the computation of the meshfree basis functions and their gradients. Additionally, RPIMC was found to be 1.65 times slower than FEM in solving the monodomain model in a large-scale biventricular model. Nevertheless, it should be noted that RPIMC is more efficient than other meshfree methods such as SPH. In  \cite{lluch2017smoothed}, the execution time to run a simulation in an SPH implementation of the monodomain model was reported to be of 57 minutes for a ventricular model with 51037 nodes and a simulation time of 150 ms, while in this study RPIMC required 99.6 minutes for a biventricular model of 273919 nodes and a simulation time of 1000 ms.

In summary, the RPIMC method is shown to be a good alternative to FEM with satisfactory accuracy for the solution of transient diffusion problems, such as heat conduction, and with good capabilities for the solution of the monodomain model in cardiac electrophysiology simulations. Importantly, RPIMC is expected to be a promising option to solve coupled electromechanical problems in cardiology, where it could outperform FEM in problems involving large displacements.

\section*{Acknowledgements}
This work was supported by the European Research Council under the grant agreement ERC-StG 638284, by Ministerio de Ciencia e Innovaci\'on (Spain) through project PID2019-105674RB-I00 and by European Social Fund (EU) and Arag\'on Government through BSICoS group (T39\_20R) and project LMP124-18. Computations were performed by the ICTS NANBIOSIS (HPC Unit at University of Zaragoza).

\bibliographystyle{unsrt}  
\bibliography{main}

\end{document}